# Characterization and modeling of thermally-induced doping contaminants in high-purity Germanium

V. Boldrini[1,2], G. Maggioni[1,2], S. Carturan[1,2], W. Raniero[2], F. Sgarbossa[1,2], R. Milazzo[1], D. R. Napoli[2], E. Napolitani[1,2], D. De Salvador[1,2]

[1] Dipartimento di Fisica e Astronomia, Università degli Studi di Padova, Via Marzolo 8, I-35131 Padova, Italy
[2] INFN-LNL, Viale dell'Università 2, I-35020 Legnaro, Padova, Italy

**Abstract**
High purity Ge (HPGe) is the key material for gamma ray detector production. Its high purity level ($\leq 2 \cdot 10^{-4}$ ppb of doping impurity) has to be preserved in the bulk during the processes needed to form the detector junctions. With the goal of improving the device performance and expanding the application fields, in this paper many alternative doping processes are evaluated, in order to verify their effect on the purity of the material. In more detail, we investigated the electrical activation of contaminating doping defects or impurities inside the bulk HPGe, induced by both conventional and non-conventional surface doping processes, such as B ion implantation, P and Ga diffusion from Spin-On Doping (SOD) sources, Sb equilibrium diffusion from a remote sputtered source and laser thermal annealing (LTA) of sputtered Sb. Doping defects, thermally-activated during high temperature annealing, were characterized through electrical measurements. A phenomenological model describing the contamination process was developed and used to analyze the diffusion parameters and possible process thermal windows. It resulted that out-of-equilibrium doping processes confined to the HPGe surface have higher possibilities to be successfully employed for the formation of thin contacts, maintaining the pristine purity of the bulk material. Among them, laser thermal annealing turned out to be the most promising.

## 1. Introduction

In the last two decades, germanium has become one of the most studied semiconductors and now it is applied in many research and applicative fields such as microelectronics, photonics, solar energy and radiation detectors. In order to improve performances, a great effort is demanded to find new doping technologies [1]. Particularly, in the field of γ-ray detectors research, the development of suitable doping processes to form a segmentable n-type contact on high purity germanium (HPGe) is crucial to improve the performance of gamma trackers with high energy resolution [2,3]. Currently, n-type gamma detectors are available with a segmented p$^+$ contact made by B ion implantation. Unfortunately, basing the gamma-ray tracking on the hole signal is not efficient because of the high density of hole traps that are formed during usage and annealing, with consequent leakage current and worsening of the energy resolution [4]. For these reasons, HPGe detectors would benefit from the segmentation of the n$^+$ contact, which nowadays is prevented by the use of Li thermal diffusion technique: the contact depth is very large (mm) and further increases after annealing, due to Li high diffusivity, causing segmentation fault. Moreover, the thick Li n$^+$ region is a dead layer with no field and this reduces the sensitivity toward low-energy gamma rays and worsens the tracking performance. The improvement of such technological issue may open the route for future medical or security devices with gamma ray directional sensitivity and high energy resolution, without the need of collimators.
Standard doping techniques include the use of high temperature annealing treatments, which could cause the electrical activation of doping defects or impurities inside the whole crystal volume. In the perspective of finding more suited doping techniques for the formation of the n-type contact in γ-ray HPGe detectors, we developed P and Ga diffusion from SOD sources [5], Sb equilibrium diffusion from a remote sputtered source [6] and laser thermal annealing (LTA) diffusion of sputtered Sb [2]. All of them require high temperature annealing treatments, hence, to preserve the bulk purity thus saving the detector operation, the issue of impurity activation should be faced.
Nowadays, germanium crystals with impurity concentration lower than $10^{10}$ cm$^{-3}$ are available. Despite growth purity, thermal processes often needed to make p-n junctions promote the diffusion of impurities and defect states throughout the material and the incorporation of external contaminants through physical processes that are thermally activated. All these species such as dislocations, shallow impurities (Cu, Ga, B, Li, etc.) and deep-level contaminants (Cu, Fe, Zn, etc.) can increase the doping level of HPGe and also act as generation/recombination centers for free charge carriers, thus affecting the operation of any device [7]. Among the aforementioned, copper is one of the fastest diffusants in Ge at low temperature [8].

In this paper, four-wire electrical measurements vs. temperature were performed on HPGe samples treated with aforementioned processes. Data about the concentration, type and mobility of charge carriers generated by thermally-activated bulk impurities were collected and modelled, in order to find a possible process window for doping HPGe without contaminating the bulk material.

## 2. Experimental

*2.1 Sample preparation*

Two (100) HPGe wafers of p- and n-type, 2 mm thick, with growth impurity concentration in the range (0.4 to 2)$\times 10^{10}$ cm$^{-3}$ and a dislocation density of about 2000 counts cm$^{-2}$, were supplied by Umicore. They were manually polished in order to smooth surfaces and then cut into 10$\times$10 mm$^2$ samples using an automatic dicing machine (Disco Corporation, Tokyo, Japan). Each sample was cleaned with hot 2-propanol, hot deionized water and HF 10% wt. to remove dicing adhesive residue and native oxides. After, a more aggressive etching bath was done in HNO$_3$(65%):HF(40%) 3:1 solution for 5 minutes, in order to remove about 100 µm and consequently the residual mechanical damage from surfaces [9].

Two p-type and n-type samples were characterized as reference samples, in order to check if the concentration of active-impurities derived from our measurement did coincide with that guaranteed by the seller.

Other HPGe slices prepared at the same way, were surface doped with different approaches. One sample was doped by B ion implantation, a standard technique used for the formation of the p-type contact in HPGe detectors. The following parameters were used for the implantation: 22.6 keV energy and 1$\times 10^{15}$ cm$^{-2}$ dose. Some samples were doped by P and Ga spin-on doping, using the experimental technique described in Ref. [5]. Commercial sources (sol-gel precursors, Filmtronics, Butler PA, USA) containing P and Ga were homogeneously deposited above the sample front surface by spin-coating. Then, the films underwent a curing stage at 130 °C for 30 min, in a N$_2$ atmosphere at 10% relative humidity. Finally, each sample was capped with Si and quartz slices and annealed inside a standard tube-chamber furnace, through a fast annealing treatment characterized by rapid sample insertion followed by a rapid heating ramp (up to 610 °C in 12 min) and a rapid extraction.

Other doping techniques that were tested are the following: Sb equilibrium diffusion in furnace from a remote source and Sb out-of-equilibrium diffusion by LTA. In the first case, a thin film of pure Sb was sputtered on an auxiliary Si substrate, which then was placed in a quartz boat over the HPGe front surface at a distance of 8.5 mm. During the annealing performed at 605 °C for 30 min, thanks to the distance set between HPGe surface and the source, Sb diffuses inside HPGe without causing surface damages [6]. In the second case, i.e. Sb out-of-equilibrium diffusion, an ultra-thin film of pure Sb is sputtered directly on the HPGe surface through the aid of a mask, in order to form four square Sb sources of 1.5 mm side at the vertexes of the square surface. Then, these four Sb sources were irradiated with a fast laser pulse of 7 ns. The laser pulse melts the first 150 nm of HPGe thus inducing Sb diffusion in liquid. The melt depth has been determined through SIMS measurements [2].

In order to investigate any possible protective action against impurity diffusion inside HPGe coming from the outside, one as-cut sample was sputtered with SiO$_2$ all around before performing a standard fast annealing treatment.

In order to disentangle the effect of the thermal treatment and that of the deposition processes, some samples were annealed with different thermal budgets with no doping source deposition and one of them was annealed inside a different furnace (that will be called F2) in order to verify if the induced impurity level was dependent on the furnace. A list of all the samples that were prepared and characterized is reported in Table 1.

As concerning thermal treatments in standard furnace, the temperature vs. time curve, T(t), including heating and cooling ramps was measured through a thermocouple. All the doping processes, except for Sb LTA, were done across the entire front surface. In order to be able to measure the bulk active-impurity concentration through four-wire measurement, the removal of these highly-doped surface layers was necessary. In fact, if highly-doped layers were not removed, during the electrical measurement the current would preferentially pass through the more conductive surface layer, thus complicating the result. The removal was done through an etching bath in HNO$_3$:HF (3:1 volume ratio) solution for 20 s, leaving only four square doped areas at the sample corners, in order to easy the electrical contact during the subsequent electrical measurement. The removed Ge thickness was about 10 µm, much more than the highly doped zone that is confined in the first 0.5 µm for all the treatments.

Then, Au/Cr square pads (1.5 mm side) acting as contacts were sputtered at the corners of all the samples. All electrical measurements have been done by using thin Cu wires bonded to Au/Cr pads through malleable indium.

*2.2 Low T electrical measurements*

Four-wire electrical measurements at low temperature were performed, according to the Van der Pauw method. The experimental set-up, shown in Fig. 1, is a variable temperature Hall-effect measurement system provided by MMR Technologies. It consists of a small vacuum chamber containing a Joule-Thomson micrometer refrigeration circuit, in which high-pressure $N_2$ (~124 bar) is injected and expanded, provided with a ceramic stage for thermal contact with the sample. In the stage there is also a resistor for the heating. The sample is bonded to a Kapton printed circuit for electrical measurements (Fig. 2), which are carried out with the use of a Keithley 2600 sourcemeter, a switch matrix, a customized acquisition software and a permanent magnet for the Hall effect. As regarding the refrigeration, the apparatus is complemented by an $N_2$ filter-dryer system, capillary tubes for $N_2$ transport and injection, a rotary and a turbomolecular pump to keep a vacuum level of about $5\times10^{-5}$ mbar inside the chamber and finally, a remotely-manageable temperature controller. With this apparatus a wide range of temperature can be investigated (77-700) K.

Hall measurements can be performed with the same apparatus, by inserting the chamber in a permanent magnetic field of 0.5 T.

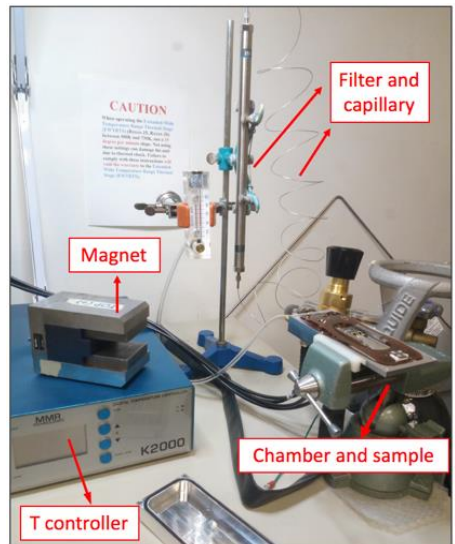

*Figure 1. Electrical measurement setup: on the right there is the open vacuum chamber containing the sample; the thin cylinder behind is the $N_2$ gas filter, connected both to the $N_2$ tank and the vacuum chamber through capillary tubes. On the left, there is the permanent magnet and the temperature controller.*

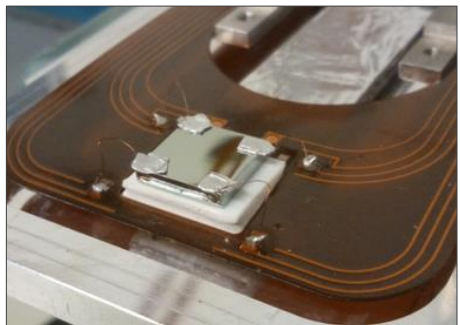

*Figure 2. Detail of a sample positioned above the ceramic part of the refrigerator circuit and soldered to the Kapton printed circuit.*

**3. Results and discussion**

*3.1 Experimental data*

Figure 3 reports the results of bulk sheet resistance measurements with decreasing temperature, performed according to the Van der Pauw method. The measurements were done with a current of 1µA, value chosen after having tested a wide range of currents looking for a plateau of sheet resistances [10]. A geometrical correction factor of 1%, including both effects of contact size and position, was estimated through finite element simulations done with COMSOL Multiphysics software and applied to data. As regarding measurement errors, due to electrical signal reproducibility we estimate a 10% error for the measurements on n-type samples (n-type Umicore and Sb LTA); in all other cases, the error is below 5%.

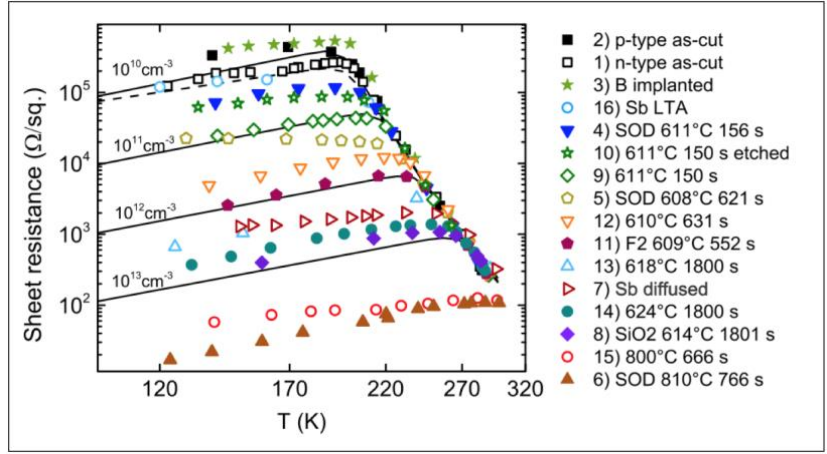

*Figure 3. Sheet resistance data as function of the measuring temperature, for all samples. Full symbols refer to p-HPGe starting substrates, empty symbols to n-HPGe. The legend reports the number and a brief description of each sample.*

Starting from room temperature, the sheet resistance rapidly increases until a maximum value is reached that is very different from sample to sample (from $10^2$ to $5\times10^5$ Ω/sq.). Then, for all the samples the sheet resistance reverses its trend and it slowly decreases by continuing lowering the temperature.

An interesting result comes out from Hall-effect measurements at low temperature, done on a set of selected samples (Table 1, last column). At low temperature, all samples that had received a high-temperature annealing process showed positive carrier sign, even if their bulk was n-type before any treatment. The only samples that showed n-type carriers were the reference n-type HPGe and the sample that received Sb doping by laser thermal annealing. This result is important because it shows that during high-temperature annealing in a standard furnace, a strong activation of acceptor levels occurs inside bulk HPGe, while laser thermal annealing technique seems not to involve this kind of problem.

In order to better understand the sheet resistance data, the expected theoretical trends of sheet resistance as function of temperature were calculated for different doping levels. By assuming a square p-type Ge sample, 1 cm$^2$ area and thickness t, the theoretical sheet resistance $R_{sheet}$ can be calculated by:

$R_{sheet} = 1/(tpe\mu_h)$ (1)

where *p* is the extrinsic hole density and $\mu_h$ is the hole mobility. In a p-type semiconductor with $N_a$ acceptor density, *p* can be expressed by:

$p = (N_a/2)+((N_a^2/4)+n_i^2)^{1/2}$ (2)

where

$n_i = (N_v N_c)^{1/2} \exp(-E_g/(2k_B T))$ (3)

$N_a$ being the extrinsic acceptor density, while $N_v$ and $N_c$ are the valence and conduction band effective density of states, respectively [11].

In order to calculate sheet resistance curves, the variation of carrier mobility with temperature for different carrier concentrations has to be considered in Eq. 1. This was extrapolated from literature data, particularly from Ref. [12, 13] for p-type and from Ref. [14, 15] for n-type. Literature mobility curves were digitalized and fitted by power law trends, then the power law coefficient was interpolated as a function of the doping concentration in order to have access to the mobility at any temperature and dopant concentration [16].

In Fig. 3 all theoretical sheet resistance curves as a function of temperature are shown for different p-type doping concentration (continuous lines). As can be observed, functions have all the same shape, but they present a different height in the plot depending on the carrier density. Each function is characterized by two slopes. Starting from room temperature (300 K), the first trend is typical of the intrinsic regime where $p = n_i$ and it exponentially increases by decreasing temperature according to Eq. 3.

At a certain temperature value, the density of charge carriers originated from acceptors ionization $N_a$ exceeds that of thermally activated carriers $n_i$ and a trend reversal occurs entering the saturation regime where $p = N_a$. The transition temperature is different for each sample because the lower the impurity concentration, the lower is the temperature at which $N_a > n_i$ holds. In saturation regime, the decrease of sheet resistance by decreasing temperature derives from carrier mobility increasing due to less lattice vibrations.

The dashed line represents the sheet resistance trend for a $1\times10^{10}$ cm$^{-3}$ n-type doped germanium and is calculated by the n-type version of Eq. 2. As can be noted the low temperature sheet resistance drop shows a different slope depending on the impurity type (p- or n-) due to different mobility dependence on temperature [11].

*3.2 Contaminant concentration analyses*

Sheet resistance experimental data vary with temperature as expected, in fact the shape of experimental curves is the same as theoretical lines. Interestingly, experimental curves lie on different heights in the plot, meaning that our samples are characterized by different bulk impurity concentrations.

Sheet resistance data can provide the carrier density by inverting Eq.1. To do this, literature mobility can be used as described before, or alternatively mobility estimates by Hall measurement could be used. In Fig. 4 a comparison between literature and Hall measurements of the mobility are reported for a set of samples and temperatures. The data present a substantial agreement in a wide range of values; discrepancy can be attributed to low accuracy in our measurements. As a matter of fact, since Hall measurements are much more sensitive to contact geometry, we estimated 15% correction due to square contacts but larger variation may be due to poor contact shape geometry definition due to indium pads. For these reasons, and due to the fact that we performed mobility measurements only in a subset of samples and temperatures, we decided to adopt literature mobilities to calculate carrier densities.

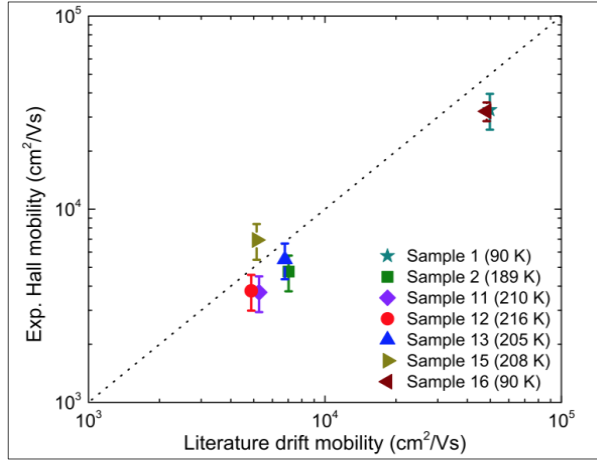

*Figure 4. Comparison between experimental Hall mobilities and literature mobilities. 20% errors are reported to the measured Hall mobility. The legend reports the sample number and the measuring temperature.*

Carrier density curves representing the density of electrically active charge carriers as function of $1/k_BT$, are shown in Fig. 5. The intrinsic and saturation regimes are clearly distinguishable and the carrier density in the saturation regime corresponds to the density of ionized bulk impurities. Hence, for each curve we took the average concentration value N in the first 50 K of the saturation regime, as the density of active ionized impurities. N values for all samples are reported in the sixth column of Table 1. Errors correspond to standard deviation of the above mentioned average.

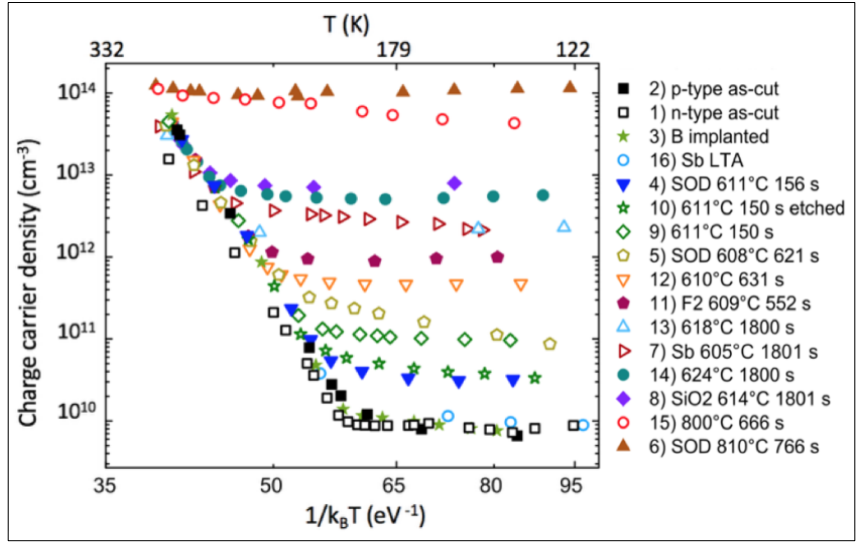

*Figure 5. Charge-carrier density curves, function of $(k_BT)^{-1}$ obtained from sheet resistance data. Full symbols refer to p-type starting substrates, empty symbols to n-type. The legend reports the number and a brief description of each sample.*

The concentration of bulk contaminants activated during our fabrication processes is not directly N since we have to consider the starting doping level of the material. In the most frequent case, an n-type sample is contaminated and becomes p-type. The amount of acceptors $N_c$ needed for this is the sum of the starting n-type dopant concentration ($N_{growth}$) to be compensated, plus N positive carriers that are measured, i.e. $N_c = N + N_{growth}$. All possible combinations of starting and final dopant concentration can be described by the formula:

$$N_c = |(+ -)N - (+ -)N_{growth}| \qquad (4)$$

using the + sign for p-type dopant and – for n type.

Growth impurity densities $N_{growth}$ are taken from the impurity level of reference samples (first two N values in Table 1), equal to $8.7 \times 10^9$ cm$^3$ for n-type and $8.9 \times 10^9$ cm$^3$ for p-type. The concentration of thermally-induced acceptor defects $N_c$ is reported in Table 1, for each sample.

*Table 1. List of all characterized samples and their most important parameters.*

| Sample number | Starting carriers | Annealing treatment | Surface treatment | Surface etching | N (cm$^{-3}$) | $N_c$ (cm$^{-3}$) | Hall final carriers |
|---|---|---|---|---|---|---|---|
| 1 | n | - | - | yes | [8.7±1.5]x10$^9$ | - | n |
| 2 | p | - | - | no | [8.9±3.3]x10$^9$ | - | p |
| 3 | p | - | B impl. | yes | [9.5±2.1]x10$^9$ | - | - |
| 4 | p | 611C 156s | P diff. | yes | [3.4±0.6]x10$^{10}$ | [2.5±0.9]x10$^{10}$ | - |
| 5 | n | 608C 621s | P diff. | yes | [2.0±1.0]x10$^{11}$ | [2.1±1.0]x10$^{11}$ | - |
| 6 | p | 810C 766s | Ga diff. | yes | [1.1±0.1]x10$^{14}$ | [1.1±0.1]x10$^{14}$ | - |
| 7 | n | 605C 1801s | Sb diff. | yes | [3.0±0.9]x10$^{12}$ | [3.0±0.9]x10$^{12}$ | - |
| 8 | p | 614C 1801s | SiO$_2$ sputt. | no | [7.8±1.0]x10$^{12}$ | [7.8±1.0]x10$^{12}$ | - |
| 9 | n | 611C 150s | - | no | [1.1±0.2]x10$^{11}$ | [1.2±0.2]x10$^{11}$ | - |
| 10 | n | 611C 150s | - | yes | [4.8±1.6]x10$^{10}$ | [5.7±1.8]x10$^{10}$ | - |
| 11 | p | F2 609C 552s | - | no | [9.9±1.5]x10$^{11}$ | [9.8±1.5]x10$^{11}$ | p |
| 12 | n | 610C 631s | - | no | [5.0±0.8]x10$^{11}$ | [5.1±0.8]x10$^{11}$ | p |
| 13 | n | 618C 1800s | - | yes | [2.2±0.3]x10$^{12}$ | [2.2±0.3]x10$^{12}$ | p |
| 14 | p | 624C 1800s | - | yes | [5.7±1.1]x10$^{12}$ | [5.7±1.1]x10$^{12}$ | - |
| 15 | n | 800C 666s | - | yes | [6.9±2.2]x10$^{13}$ | [6.9±2.2]x10$^{13}$ | p |
| 16 | n | LTA | Sb sputt. | no | [1.1±0.2]x10$^{10}$ | - | n |

Looking at the $N_c$ column in detail, it can be seen that B ion implantation is a clean process as expected, since it does not introduce any further contaminant into HPGe. A really remarkable result is obtained with laser thermal annealing of Sb: its impurity level is fully compatible with the starting material level. This means that this doping technique does not introduce shallow levels inside HPGe, as also confirmed by the measurement of the carrier type, that has not changed to p as it happened for other processes.

Regarding the $SiO_2$ coating, it has not provided any protective action since the underlying bulk is contaminated even more than a naked sample that had received the same annealing treatment. Samples coated with SOD sources are characterized by the same (or even less) impurity density as other as-cut samples that received a similar annealing treatment. This means that the SOD film does not introduce further impurities inside Ge. The same is true for the sample doped with Sb from a remote source. It is interesting to look at the sample treated inside a different furnace (F2), because it turns out to have a slightly higher impurity density than the others annealed with the same temperature ramp. This could suggest that active doping species under study are not intrinsically present inside HPGe or formed during the process and then activated during the annealing, but probably they are impurities coming from the external environment and diffusing inside HPGe during the thermal treatment. Anyway, the most evident behavior emerging from the observation of the table is that the concentration of doping defects increases by moving to higher thermal budgets.

*3.3 Phenomenological model of contamination process*

The contamination of the HPGe crystal may occur because of the diffusion of external contaminants, or because of the formation of intrinsic defects under annealing (dislocation multiplication, point defects clustering, etc.).

A phenomenological model has been created that allows to explain the contamination process under furnace annealing. In a first step we considered the pure thermal effect as the only cause of contamination. In a second step we will try to understand if there are also some other phenomena that may influence the contamination. The phenomenological evaluation through such a model will provide a useful framework to explore the possibility of finding doping process windows while keeping the contamination under an acceptable limit.

The first hypothesis of the model is a standard Arrhenius relation between the equilibrium impurity/defect concentration achieved with long annealing treatments, $n_{eq}$, and temperature:

$$n_{eq} = n_0 \exp(-E_{act}/(k_B T)) \qquad (5)$$

where $n_0$ is the concentration pre-factor and $E_{act}$ is the activation energy.

The second hypothesis is a first order non-equilibrium dynamics, according to which the growth rate of dopant contaminants is proportional to their "distance" from the equilibrium:

$$dn/dt = r (n_{eq} - n) \qquad (6)$$

where $n$ is the actual amount of contaminant and $r$ is the contamination rate constant, which fixes the velocity by which the equilibrium is restored. We assume that the furnace annealing treatments are short (8-30 min) and therefore we make the reasonable assumption that the system is always far from the equilibrium i.e. $n_{eq} \gg n$. Given such approximation, and substituting (5) into (6) we get:

$$dn/dt = r\, n_0 \exp(-E_{act}/(k_B T(t))) \qquad (7)$$

where the dependence of T on time is explicitly expressed. The integration of Eq. 7 over time returns:

$$n = r\, n_0 \int \exp(-E_{act}/(k_B T(t)))\, dt \qquad (8)$$

Eq. 8 can be used to model the contamination data $N_c$ as a function of the thermal treatment ramp $T(t)$, once given the parameters $r \cdot n_0$ and $E_{act}$. In order to facilitate the model calibration, we introduce the quantity TB (thermal budget) as the integral term in Eq. 8:

$$TB = \int \exp(-E_{act}/(k_B T(t)))\, dt \qquad (9)$$

TB can be computed for each process (once the activation energy has been fixed) by using the function T(t) that was registered through a thermocouple during each furnace treatment. It is worth to note that such number summarizes the effect of a thermally activated process during a general thermal history, including heating and cooling ramps that may be significant for short annealing times.

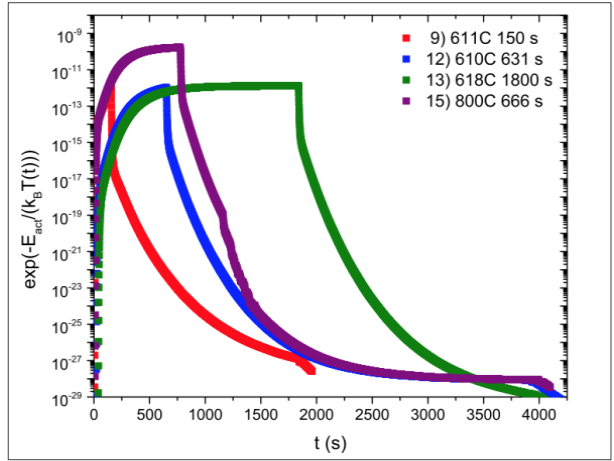

*Figure 6. Arrhenius function of all fast-annealing treatments, where temperature depends on time, calculated for the same activation energy of 2.1 eV. The legend reports the sample number and the annealing treatment.*

In Fig. 6 the plot of $\exp(-E_{act}/(k_B T(t)))$, for one peculiar value of activation energy (2.1 eV), is shown for all fast-annealing treatments. Thermal budgets are calculated as the integral of these curves.
In order to facilitate the fitting and to take into account the large dynamics of the contamination data, we apply the natural logarithm to equation 8, getting a linear relation between *ln(n)* and *ln(TB)*:

$$ln(n) = ln(r \cdot n_0) + ln(TB) \qquad (10)$$

This last equation can be used to fit our experimental data, particularly those done in the same furnace, using $E_{act}$ and $r \cdot n_0$ as free parameters. We tested possible values of $E_{act}$ in the range from 0.6 to 3 eV. Once calculated the thermal budgets for all samples, it was possible to create a plot *ln(n) vs ln(TB)*, where each point refers to a single sample. Points should lie on a straight line with a slope equal to 1, according to Eq. 10.

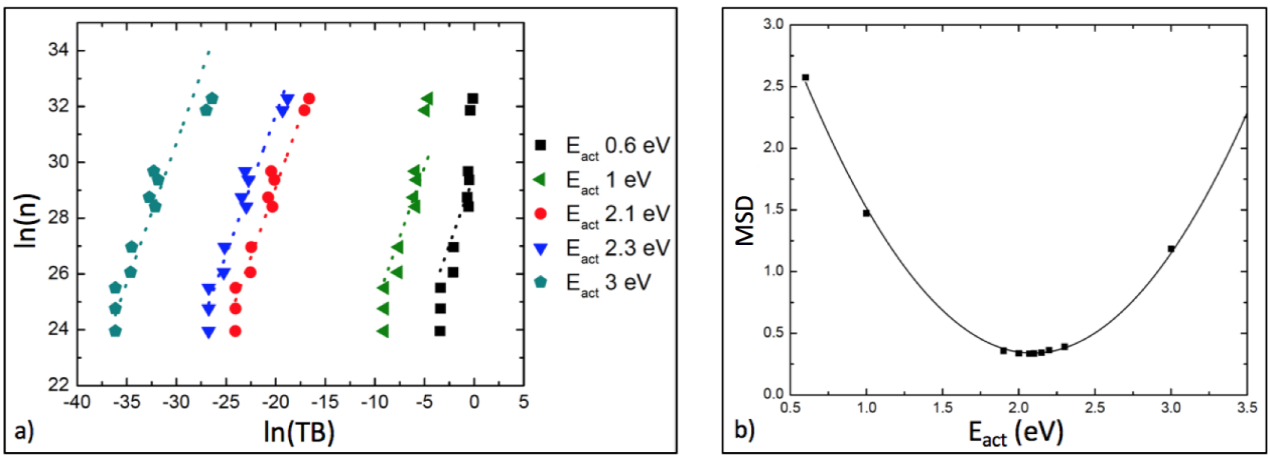

*Figure 7. (a) Data are reported several times, using different values of activation energy in the calculation of thermal budgets. Dashed lines represent linear fits with slope fixed to 1. (b) For each tested value of activation energy, the mean square deviation (MSD) of data from each fitting is reported.*

In Fig. 7(a), groups of points corresponding to different values of activation energy are reported. Each group was fitted with a linear function, by keeping fixed the slope to 1. In Fig. 7(b), the mean square deviation (MSD) between the data and the fit is plotted for the different activation energy. The minimum MSD determines the best possible value for the activation energy $E_{act} = (2.1 \pm 0.1)\ eV$. The corresponding $r \cdot n_0$ value is $2.1 \times 10^{21}\ cm^{-3}\ s^{-1}$.

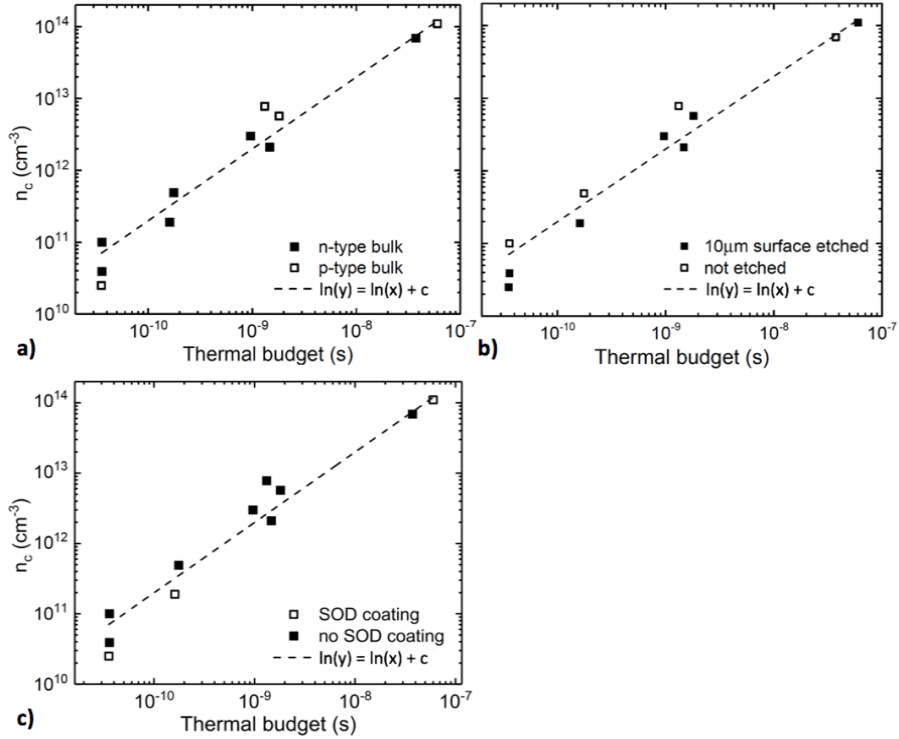

*Figure 8. Best linear fit of the data. (a) Comparison between samples with starting n-type bulk and starting p-type bulk. (b) Comparison between samples whose surface was chemically etched (~10 μm) and entire samples. (c) Comparison between SOD coated samples and all the others.*

In Fig. 8, the best fit is reported, i.e. with 2.1 eV activation energy. As can be noted, the model with only two free parameters allows to order the data in a well-defined trend. The mean square relative displacement of each data from the fit is 34%. This is a very good result considering that the data set comes from very different annealing procedures: temperatures from 600 to 800 °C, times from 2 to 30 minutes and different shapes of the temperature ramp. We can for sure state that the thermal budget is the main parameter causing the contamination.

In order to understand if the average displacement hides some systematic dependence on other experimental parameters, we divided the data in different subsets and inspected the trend with respect to the average model. In Fig. 8(a) we divided data according to p or n starting bulk. This could be in principle an important parameter since p and n specimens are generally taken from different zones of the ingot and may have in principle different grown-in defects. Both starting bulk types seem to be homogeneously distributed around the fitting line. This suggests that the contamination process is independent of the starting substrate type. In Fig. 8(b) we divided the sample in etched and not etched surface. It can be noted that there is a clear correlation between the impurity density and surface removal, namely etched samples present a lower impurity level. This indicates that the first micrometers of material are more contaminated and confirms that contaminants come from the external environment and diffuse inside Ge during high temperature annealing. Thus, most likely they are extrinsic, in-diffusing contaminants. This excludes all those species that, already present inside bulk HPGe, could be electrically activated during the annealing by forming complexes with other impurities or defects [7]. In Fig. 8(c) SOD-coated and not-coated samples are divided. This is in principle a fundamental step since we would like to answer the question if SOD is responsible for contamination. As a matter of fact, samples with SOD have a lower contamination density. This can hardly be attributed to a protective effect of SOD, since the $SiO_2$ coating has not proved to be an effective barrier toward contaminant diffusion. More likely, the lower contamination density can be attributed to the fact that SOD-covered samples are all surface etched. Thus, since there is not a direct correlation between the presence of the SOD source during the annealing and the contamination process, the remarkable conclusion is that the SOD film does not introduce further contaminants.

By considering the p nature of the shallow dopant and also its coming from the external environment, it could be identified with copper. Several proofs point to this identification: Cu is one of the fastest diffusing impurities in Ge. Moreover, substitutional Cu atoms give rise to three acceptor levels in Ge and this is compatible with the p nature of active impurities. According to H. Bracht et al. [8], Cu diffusion and solubility in highly dislocated Ge (as our samples are, about 2000 cm$^{-2}$) is given by::

$$D_{Cu}^{eff} = (7.8 \times 10^{-4}) \exp(-0.084 \text{ eV}/(k_BT)) \text{ cm}^2 \text{ s}^{-1}. \quad (12)$$

$$C_{Cu}^{eq} = (3.44 \times 10^{23}) \exp(-1.56 \text{ eV}/(k_BT)) \text{ cm}^{-3} \quad (13)$$

The two activation energies needed for Cu atoms to diffuse inside Ge and reach the substitutional solubility are, respectively, 0.084 eV and 1.56 eV. In order to contaminate the material, Cu has to be both solubilized and diffused into the Ge matrix. Therefore, we could expect that the activation energy of the contamination process should be at least the sum of the two energy costs, i.e. 1.64 eV. Our experimental data analysis reports a higher activation energy for contamination that is 2.1 eV. This is not in perfect agreement with our estimate, but we have to consider that, beside diffusion and bulk solubility, the surface can furnish a further barrier for the contaminant to enter the bulk or, alternatively, if the Cu availability comes from out-diffusion from the furnace walls, the activation energy of such process should be added.

As a partial conclusion, it is worth to stress that a direct identification of the doping species is not a trivial task since, also in the case of the sample with higher contamination, we are dealing with $10^{13}$ doping center/cm$^2$ i.e. a small fraction of a single monolayer.

*3.4 Process window for standard doping annealing*

In the perspective to apply new techniques for the formation of the n-type contact on HPGe, it is interesting to understand if it is possible to make the surface diffusion doping process dominate over the contamination one. For this purpose, we analyzed the diffusion process of P emitted by SOD and Sb diffusion from a remote source starting from diffusion profiles already published in Ref. [5] and [6].

We performed an analysis in term of thermal budget in order to have a comparable description of the contamination and doping processes. Diffusion lengths L were evaluated as the profile depth at half maximum. L is connected to the diffusion coefficient D by the relation:

$$L^2 = D \cdot t \quad (14)$$

Diffusion coefficient is related to the temperature by an Arrhenius relation:

$$D = D_0 \exp(-E_{act}/(k_BT)) \quad (15)$$

Therefore, substituting (15) into (14) we get:

$$L^2 = D_0 \exp(-E_{act}/(k_BT)) \, t \quad (16)$$

If the temperature T varies during the process, i.e. it is a function of time (as it happens in fast treatments) Eq. 16 can be directly generalized as follows:

$$L^2 = D_0 \int \exp(-E_{act}/(k_BT(t))) \, dt. \quad (17)$$

The last integral is defined as the thermal budget of the annealing treatment. By applying the natural logarithm to the equation, we achieve a linear relation:

$$\ln(L^2) = \ln(D_0) + \ln(TB). \quad (18)$$

Eq. (18) was used to fit diffusion length data in the same way as eq. 10 was used for bulk contamination process. In Fig. 9 we report the best fit result for P and Sb. In case of P diffusion, the best fitting was obtained with an activation energy of (2.4±0.1) eV, and $D_0 = 113.7\ cm^2s^{-1}$. In case of Sb diffusion, the activation energy is (2.7±0.1) eV, and $D_0 = 148.6\ cm^2s^{-1}$.

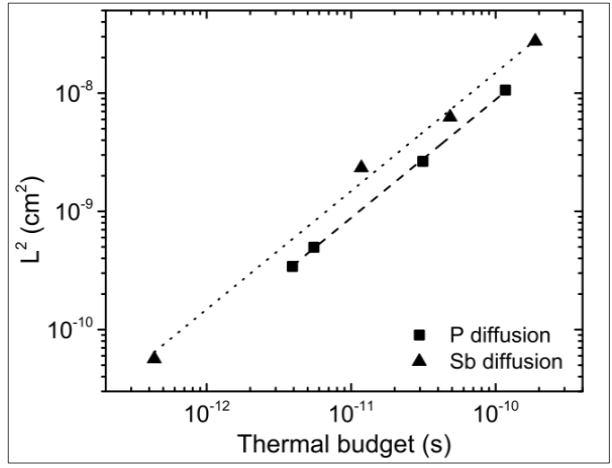

*Figure 9. Best fit of P and Sb diffusion data, obtained for $E_{act}^P = 2.4\ eV$ and $E_{act}^{Sb} = 2.7\ eV$.*

On the basis of this result it is possible to identify, if it exists, an optimal range for thermal budgets, which allows dopant diffusion inside HPGe without contaminating. We fixed an upper acceptable limit for contamination of $n_c^{thr} = 2\times10^{10}\ cm^{-3}$. According to Eqs. 8 and 9, in order to respect this threshold, TB should be limited by:

TB < ($n_c^{thr}$ /(r $n_0$))     (19a)

For simplicity, if we consider a step-like treatment, with T temperature for a time t (no ramps), Eq. 19a becomes:

t exp(-$E_{act}$/($k_B$T)) ≤ ($n_c^{thr}$ /(r $n_0$))     (19b)

By rearranging Eq. 19b, we get the maximum acceptable annealing time at a given temperature to have a contamination lower than $n_c^{thr}$ :

t ≤ $n_c^{thr}$ / ((r $n_0$) exp($E_{act}$/($k_B$T)))     (20)

In other words, by applying the natural logarithm to Eq. 20, it defines a line in the *t vs 1/($k_B$T)* space that separates non-contaminating thermal budgets from contaminating ones (continuous line in Fig. 10). A similar reasoning can be done to impose a minimum P and Sb diffusion threshold into HPGe starting from Eq. 17. Particularly, a threshold $L_{thr}$ = 200 nm for dopant diffusion length is set, in order to ensure the formation of a continuous and homogeneously-doped contact layer. We easily obtain a formula for the minimum time needed to have such doping depth:

t > $L_{thr}^2$ / ($D_0$ exp($E_{act}$/($k_B$T)))     (21)

By applying the natural logarithm to Eq. 21, it defines a line in the *t vs. 1/($k_B$T)* space that separates sufficient thermal budgets from insufficient ones to obtain a 200 nm thick doped layer. In Fig. 10 the dashed line corresponds to P diffusion threshold, while the dotted one corresponds to Sb diffusion threshold.

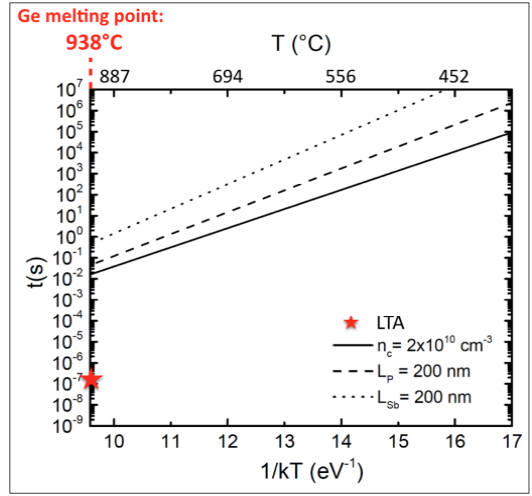

*Figure 10. Plot reporting the annealing time as function of $(k_BT)^{-1}$. The three lines represent the thresholds for the diffusion of P, Sb and contaminants (dashed, dotted and continuous lines respectively). The red star indicates the thermal budget provided by a laser annealing treatment.*

As can be observed, within the time and temperature range reachable with a standard tube furnace, it is not possible to find an optimal thermal window. In fact, a thick P- or Sb-doped layer cannot be obtained without contaminating the bulk HPGe, since lines do not cross. Anyway, by working with shorter times thus moving more and more towards a state of out-of-equilibrium, the onset of large thermal gradients within the material could allow to achieve good dopant diffusion while keeping the impurity density below the established threshold. Thus, a more rapid annealing technique that operates in the range of milliseconds, such as flash lamp annealing (FLA) [17], could be more appropriate for these purposes. The situation would change completely by moving to Ge melting temperature (938 °C), which can be done through laser thermal annealing technique. In melted Ge, dopants diffuse orders of magnitude faster than in solid, therefore the solid-state diffusion expressed by Sb and P diffusion lines completely change. Besides, this technique acts just on the first hundreds of nanometers of material, leaving completely unheated the bulk. In this way, the process of bulk contamination should be almost null. We have had a first evidence of this by characterizing the n-type HPGe sample that had received LTA of sputtered Sb. The functionality of such process to build working HPGe diodes is demonstrated in Ref. [2].

## 4. Conclusions

In this work the results of the characterization of the amount of electrically active defects found inside HPGe, after high temperature annealing treatments in standard furnaces, have been presented. Through electrical measurements at low temperature, we measured the density of these doping defects and their sign. Their density turned out to be higher when high thermal budgets were applied and the shallow levels introduced were of acceptor type. The phenomenon has been studied in samples that had received different annealing treatments, different surface doping processes and also different surface treatments. Taking into account the electrical features of these active defects and the fact that they have a higher concentration in the first 10 µm of the surface, we concluded that most likely we are dealing with copper atoms coming from the external environment.

It is worth to note that both SOD and Sb doping can be exploited by keeping a level of contamination less than $10^{12}$ cm$^{-3}$. While this level is not suitable for γ- ray detectors that work at low temperature, it is clearly good for all other room temperature applications: since thermal carriers in Ge at 300 K are $2\times10^{13}$ cm$^{-3}$, any doping below such value cannot modify the carrier concentration of devices. On the other hand, bulk contaminant behavior as recombination traps should be evaluated.

After having conceived an empirical model for the dependence of impurity density on the applied thermal budget, it was possible to fit data and determine an activation energy of (2.1±0.1) eV for the diffusion process. Through fitting of diffusion data, activation energies of (2.4±0.1) eV and (2.7±0.1) eV were also found, respectively, for P and Sb diffusion. Then it was possible to set thresholds for minimum P and Sb diffusion and maximum bulk contamination, in order to find a possible thermal window for non-contaminating doping processes. However, in the range of temperatures and times reachable with an equilibrium annealing technique such a window of allowed thermal budgets does not exist. It is evident from our results that this optimal window should be sought for shorter times, entering a regime of strongly out-of-equilibrium thermal processes and, what is most important, annealing should be limited to the Ge surface, keeping the bulk as cold as possible. Flash lamp annealing (FLA) and laser thermal annealing (LTA) could be suitable techniques to reach these conditions. A first evidence of LTA effectiveness has already been demonstrated in this work and this approach turns out to be the most promising and concrete perspective to renew and improve doping processes in HPGe materials [2].

**Acknowledgements**

This project has received funding from the INFN, 3rd Commission, GAMMA experiment and from the European Union's Horizon 2020 research and innovation programme under grant agreement n. 654002. PhD Riccardo Camattari is acknowledged for his technical support.